\documentclass[sigconf]{acmart}
\pdfoutput=1

\usepackage{amsmath}
\usepackage{amssymb}
\usepackage{amsfonts}

\usepackage{enumitem}
\usepackage{graphicx}
\usepackage{listings}
\usepackage{multirow}
\usepackage{url}
\usepackage{algorithm}
\usepackage{algorithmicx}
\usepackage[noend]{algpseudocode}
\usepackage{algpseudocode}
\makeatletter
\renewcommand{\ALG@beginalgorithmic}{\small}
\makeatother

\renewcommand\[{\begin{dmath}}
\renewcommand\]{\end{dmath}}
\newlength\myindent
\newcommand{\tech}{\textit{DroidSpan}}
\newcommand{\base}{\textit{MamaDroid}}

\def\denseitems{
  \itemsep1pt plus1pt minus1pt
  \parsep0pt plus0pt
  \parskip0pt\topsep0pt}

\usepackage{parcolumns}
\usepackage{pifont,calc}
\definecolor{verylightgray}{gray}{0.85}

\usepackage[none]{hyphenat}
\emergencystretch=\maxdimen
\hyphenchar\font=-1

\usepackage{hyperref}

\usepackage{colortbl}

\usepackage{balance}

\definecolor{Gray}{gray}{0.85}
\definecolor{LightCyan}{rgb}{0.88,1,1}

\newcolumntype{g}{>{\columncolor{Gray}}c}
\newcolumntype{w}{>{\columncolor{white}}c}

\newfont{\ttlsc}{phvb8t at 15pt}
\newfont{\autt}{pcrr8t at 11pt}

\acmDOI{10.475/123_4}

\acmISBN{123-4567-24-567/08/06}

\acmConference[Tech. Report]{arxiv.org}{July 2018}{}

\begin{document}
%
%
%
%


%

\title{A Preliminary Study On the Sustainability of Android Malware Detection}

\author{Haipeng Cai}
\affiliation{%
  \institution{School of Electrical Engineering and Computer Science\\
Washington State University}
  \city{Pullman}
  \state{Washington}
}
\email{haipeng.cai@wsu.edu}


\begin{abstract}
Machine learning-based malware detection dominates current
security defense approaches for Android apps. 
However, due to the evolution of Android platforms and malware, 
existing such techniques are widely limited by their need for constant 
retraining that are costly, and reliance on new malware samples that may not be timely available. 
As a result, new and emerging malware slips through, as seen from the continued 
surging of malware in the wild. 
Thus, a more practical detector needs not only to be accurate but, 
more critically, to be able to {\em sustain} its capabilities over time without frequent retraining. 

In this paper, we study how Android apps evolve as a population over time, in terms of 
their behaviors related to accesses to sensitive information and operations. 
We first perform a {\em longitudinal} characterization of 6K benign and malicious apps developed across 
seven years, with focus on these sensitive accesses in app executions. 
Our study reveals, during the long evolution, a {\em consistent}, clear differentiation between 
malware and benign apps regarding such accesses, measured by relative statistics of relevant method calls.
Following these findings, we developed {\tech}, a novel classification 
system based on a new behavioral profile for Android apps.
Through an extensive evaluation, we showed that {\tech} can not only effectively 
detect malware but sustain high detection accuracy (93\% F1 measure) for four years (with 81\% F1 for five years). 
Through a dedicated study, we also showed its resiliency to sophisticated evasion schemes. 
By comparing to a state-of-the-art malware detector, we demonstrated the largely superior sustainability of 
our approach at reasonable costs. 
\end{abstract}

\maketitle

%
%
%

%
%

%
%




\section{Introduction}\label{sec:intro}
The increasing dominance of Android among mobile computing platforms~\cite{idcandroid16} is accompanied by
its share of 
the vast majority (over 90\%) of all mobile malware~\cite{MalwareStat2015}.
In response to the continued surge of malware in different markets of Android
applications (known as {\em apps}), there has been a growing body of defense
solutions against malware~\cite{tan2015securing,faruki2015android} being proposed.
A major defense technique has been app {\em classification} based on machine learning (ML), which identifies malware by predicting a given app as benign or malicious~\cite{grace2012riskranker,burguera2011crowdroid,kolbitsch2009effective,shabtai2012andromaly,galal2015behavior}.

Typically a ML-based classification approach works by first training a classifier based on
a set of features extracted from labeled sample apps, and then applying the trained classifier to
unlabeled apps using the same feature set.
Thus, the key step of such approaches is the extraction of app features that constitute a {\em behavior profile} for apps.
Existing approaches have explored various kinds of features, computed through
static~\cite{yang2014droidminer,zhang2014semantics,Arp14,avdiienko2015mining,suarez2017droidsieve},
dynamic~\cite{afonso2015identifying,dash2016droidscribe}, or hybrid~\cite{saracino2016madam,Chen2016SSM}
app analysis.
Mostly, the features are based on apps usage of permissions~\cite{saracino2016madam,suarez2017droidsieve,Chen2016SSM}
and/or APIs~\cite{wu2012droidmat,yang2014droidminer,Aafer13,Arp14,zhang2014semantics,avdiienko2015mining}.
However, due to the evolution of attack strategies of Android malware~\cite{tam2017evolution}
classifiers built on these features
may not be {\em sustainable}---they would need to be retrained constantly for later use, or their performance would
downgrade immensely. Retraining a learning model often implies feature computation on new samples, which dominates the
total cost of classification. More critically, new samples (malicious ones in particular) may not be always available for retraining
the classifier to detect {\em emerging} malware.

A most recent technique, {\base}~\cite{mariconti2017mamadroid} detects malware based on abstracted API calls.
It achieved high accuracy (up to 99\%) when the training and testing apps were developed in the same period of time,
as many previous approaches did.
The main advantage of {\base} lies in its avoidance of continuous retraining: it can detect malware appeared ($N$ years)
after training, and thus attained the state-of-the-art sustainability.
Yet, {\base} kept reasonably good performance for only one year.
Its detection accuracy dropped noticeably over time, to 75\% for $N=2$ and below 50\% for $N\ge3$.

In this paper, we propose a new behavioral profile for Android apps, called {\em sensitive access distribution (SAD)},
that models their run-time behaviors by capturing quickly-exposed patterns of accesses to sensitive data and
operations through short tracing.
Given an app, the SAD profile characterizes its execution in terms of the extent and distribution
of invocations of sources and sinks, and those of sensitive control flows at method level.
Each SAD profile is described by 52 dynamic features, computed only from the trace of {\em all} method calls
in the app that are exercised during the tracing. Thus, the profile construction does not need static analysis
but only straightforward, lightweight bytecode instrumentation.
With respect to this dynamic profile, we have performed a {\em longitudinal} characterization of 6K benign and malicious
apps collected across the past seven years, in order to understand their behavioral evolution.
Based on the SAD profile, we also have developed {\tech}, a novel system for malware detection 
aiming at 
{\em superior sustainability} over existing approaches.

Our characterization reveals that sensitive accesses are prevalently {\em executed} in all apps
over all the examined years, with expected lower extent and frequency of use in benign apps than in malware.
Also, in terms of the functionalities of these accesses, network information and account settings are the
top dominating data and operations accessed, respectively, in both benign and malicious apps (but again with
less extent of use in benign apps).
From an evolution perspective, benign apps tend to be much more stable than malware in terms of the whole
SAD profile. Most notably, despite the fluctuation in both malware and benign and evolution of the Android system itself,
there was a clear and {\em consistent} separation in this profile between the these two groups over the years.

We have extensively evaluated the performance of {\tech} 
against 5K real-world apps, with both cross validation and independent testing (for new/unseen samples).
Our evaluation shows that {\tech}
achieves the level of classification accuracy comparable or superior to in contrast to existing systems
such as {\base}, when testing apps developed in the same time period as the training data.
Notably, {\tech} maintained high accuracy (F-measure of over 93\%) after as long as four years, and
still kept a reasonable performance (F-measure of 82\%) even five years after training.
We further demonstrated the resilience of {\tech} to malware evasion schemes by evaluating it on a benchmark suite of
obfuscated apps, on which {\tech} sustained the same level of classification capabilities.
With the same dataset and evaluation settings,
{\base} sustained over 90\% accuracy for one year only while not able to work on any of the obfuscated benchmarks.
Notably, our characterization findings essentially explained the sustaining capabilities of our technique. 

The main contributions of this work include:
\begin{itemize}[leftmargin=14pt]
\denseitems
\item A longitudinal characterization of run-time sensitive accesses by Android apps spanning seven years, which sheds new light
    on the evolutionary patterns of application security behaviors in Android.
\item A novel malware detection approach {\tech} based on a new behavioral profile,
    {\em sensitive access distribution}, for Android apps defined by a small feature set, which captures quickly-exposed sensitive access patterns in shortly traced app executions.
\item An extensive evaluation of {\tech}, which shows
    its comparable classification performance for same-period detection, but greatly superior sustainability and resiliency in contrast to the state-of-the-art malware detection system, at reasonable costs.
\item An open-source implementation of {\tech} and shared experimental datasets (link withheld), which facilitate replication of this work and development of future works on app security.
\end{itemize}

\section{Problem and Motivation}
ML-based classification has been a major approach to
Android malware detection. This approach also has been
successful with high accuracy, yet mostly only
for classifying apps developed in the same time period
as the training data. The reason is that both the
attack strategies of malware and the platforms of Android itself
evolve rapidly. As a result, the classifier needs to be constantly
retrained with new malware samples (i.e., they are not sustainable).
Developing a sustainable malware detector is crucial for Android
because otherwise the detector, without retraining, would not be able to identify
new malware, which presently continues to emerge and surge.
Even if retraining is an option, it is not always practical because getting
training samples for emerging malware may not be possible.
Meanwhile, the cost of computing features and training typically 
dominates the total detection cost.
Systems like MudFlow~\cite{avdiienko2015mining} did not rely on malware samples but suffered from relatively
low detection accuracy (86\%) and high overheads while their sustainability was not clear.
The state-of-the-art malware detector with demonstrated sustainability, MamaDroid~\cite{mariconti2017mamadroid},
made important advance in this regard, yet it only sustained high accuracy for one year.
Our experience showed that retraining MamaDroid is highly expensive in both time and storage.

We thus set out to first understand how Android apps evolve over a long period of time in a longitudinal characterization
study. We then leverage the findings from the study to develop a sustainable malware detector. We show how and why our approach
works, both technically and empirically.

\section{Dynamic SAD Profile}\label{sec:bkg}
Underlying both our characterization study and {\tech} is the dynamic SAD profile,
a new behavioral model of Android apps that characterizes the run-time accesses
to sensitive data (by invoking sources) and operations (by invoking sinks).
Next, we describe the defining metrics (i.e., features) of a SAD profile and how
the profile is constructed for an app.

\subsection{SAD Profile Definition}
The SAD profile of an app is defined by 52 features, much simpler than
the app profiles used by most existing ML-based app classifiers (which often use
tens of thousands of features, e.g., 150K in the best-performing mode of {\base}).
The central focus of all these features is on the distribution of source/sink invocations, albeit
they fall in three classes each addressing a different perspective: (1) the overall {\em extent} of
sensitive access in terms of total source/sink invocations, (2) the {\em categorization} of
invoked sensitive accesses in terms of the categories of data retrieved by the calls to sources, and
the categories of operations performed by the calls to sinks, and (3) the method-level control {\em flows}
that potentially carry out execution paths reaching to sink calls from source calls.

%
%
\setlength{\tabcolsep}{1pt}
\begin{table}[!h]
\vspace{-0pt}
  \centering
  \caption{Features (52) that constitue a SAD profile}\vspace{-10pt}
    \begin{tabular}{p{2.5cm}|p{6cm}}
    \hline
    \textbf{Features} & \textbf{Description (percentage of)} \\
    \hline
    \multicolumn{2}{l}{\textit{\color{blue}{Extent of sensitive accesses}}} \\
    \hline
    \hline
    Total source/sink callsites (2) & callsites targeting sources (resp. sinks) over all callsites. \\
    \hline
    Total source/sink call instances (2) & instances of calls to sources (resp. sinks) over all call instances. \\
    \hline
    \multicolumn{2}{l}{\textit{\color{blue}{Categorization of sensitive data and operations accessed}}}  \\
    \hline
    \hline
    Sensitive callsite distribution (11) & callsites targeting sources (resp. sinks) that are in each category (out of 5, resp. 6) over all source (resp. sink) callsites. \\
    \hline
    Distribution of sensitive call instances  (11) & instances of calls to sources (resp. sinks) that are in each category (out of 5, resp. 6) over all source (resp. sink) call instances. \\
    \hline
    \multicolumn{2}{l}{\textit{\color{blue}{Sensitive (potentially vulnerable) method-level control flows}}} \\
    \hline
    \hline
    Vulnerable source/sink callsites (2) & callsites targeting vulnerable sources (resp. vulnerable sinks) over source (resp. sink) callsites. \\
    \hline
    Vulnerable source/sink call instances (2) & instances of calls to vulnerable sources (resp. vulnerable sinks) over all source (resp. sink) call instances. \\
    \hline
    Vulnerable callsite distribution (11) & callsites targeting vulnerable sources (resp. vulnerable sinks) that are in each category (out of 5, resp. 6) over all vulnerable source (resp. sink) callsites. \\
    \hline
    Distribution of vulnerable call instances (11) & instances of calls to vulnerable sources (resp. vulnerable sinks) that are in each category (out of 5, resp. 6) over all source (resp. sink) call instances. \\
    \hline
    \end{tabular}%
  \label{tab:features}%
  \vspace{-5pt}
\end{table}%

Table~\ref{tab:features} gives a summary definition of the features that constitute a SAD profile, with
the sizes (in parentheses) of each feature subset. All features are consistently percentages of calls
in one group over those in total. We use these {\em relative} statistics instead of sheer numbers in order to
capture the {\em general patterns} of app behaviors in sensitive accesses.
We examine each app trace in two complementary views: {\em callsite} view concerning unique callsites regardless of
the times a callsite is invoked, and {\em instance} view counting all call instances.
For each feature in the callsite view, there is a {\em dual} feature in the instance view.

An app may invoke sources and sinks for legitimate purposes. Thus, source/sink invocations can only
be regarded as {\em sensitive} but not necessarily {\em vulnerable}.
We consider a source {\em vulnerable} if it reaches at least one sink callsite, and a sink {\em vulnerable} if there is
at least one invoked source that reaches the sink, both through method-level control flows
(i.e., dynamic call sequences).
While a conservative (imprecise) approximation, the reachability based on control flows at method level
is cheaper to compute and potentially more resilient to malware evasions~\cite{maiorca2015stealth}, compared to tracking sensitive
data flows~\cite{avdiienko2015mining}. This is also supported by our evaluation results with obfuscated benchmarks.
Note that we do not count the numbers of sensitive flows (nor compute all of them)
but only use the reachability in the SAD profile.

We categorize sensitive data accesses with respect to five categories of information sources
mostly retrieve (i.e., {\em Account}, {\em Calendar}, {\em Location}, {\em Network info}, and {\em System configurations}).
We also categorize sensitive operations with respect to six categories of operations sinks mostly perform (i.e., {\em Account setting}, {\em File operation}, {\em Logging}, {\em Network access}, {\em Messaging}, and {\em System setting}).
We chose these categorizes because they were the predominant ones according to our previous empirical study
in this regard~\cite{cai2016understanding,Cai2017androidstudy}.

\subsection{SAD Profile Construction}
Constructing the SAD profile for an app is reduced to extracting the 52 features from the app.
We use our Android app characterization toolkit~\cite{Cai2017droidfax,cai2017artifacts} 
based on Soot~\cite{lam11oct} for the extraction.
First, the app is instrumented for tracing. To overcome the vulnerability of static analysis to
code obfuscation, we probe for invocation of all methods in the APK bytecode, including those in exception-handling
constructs (e.g., {\tt catch} and {\tt finally} blocks) and those invoked via reflection.
Next, the instrumented app is exercised with automatically generated test inputs for $T$ minutes.
Lastly, all the features are computed from the collected trace.
To identify the calls to sources and sinks from the trace, we use the source/sink lists
generated by SUSI~\cite{rasthofer2014machine} and our manual categorization 
that refines the original one~\cite{Cai2017androidstudy,cai2017artifacts}.
A dynamic call graph with call-frequency annotations is built to facilitate computing the features, especially
those based on the statistics about vulnerable sources and sinks.
Note that since we use very large and conservative lists of
predefined (18K) sources and (7.2K) sinks, we consider accesses in apps sensitive very conservatively (which
is reasonable for detecting malware).


\section{Datasets \& Characterization}
\label{sec:study}

\setlength{\tabcolsep}{3pt}
\begin{table}[tp]
  \centering
  \caption{Overview of our datasets}
    \begin{tabular}{|r|r|r|r|r|r|}
    \hline
    \textbf{Category} & \textbf{Name} & \textbf{Year(s)} & \textbf{\#A.S.} & \textbf{\#C.S.} & \textbf{\#E.S.} \\
    \hline
    \multicolumn{1}{|c|}{\multirow{3}[6]{*}{\shortstack{Benign\\
apps}}} & \textit{oldBen} & 2010-2014 & 1,221 & 1,111 & 1063 \\
\cline{2-6}    \multicolumn{1}{|c|}{} & \textit{newBen} & 2017 & 2,210 & 2,048 & 1612 \\
\cline{2-6}    \multicolumn{1}{|c|}{} & \textit{\color{blue}{total}} & \multicolumn{1}{r|}{\textit{\color{blue}{2010-2017}}} & \multicolumn{1}{r|}{\textit{\color{blue}{3,431}}} & \multicolumn{1}{r|}{\textit{\color{blue}{3,159}}} & \textit{\color{blue}{2,675}} \\
    \hline
    \multicolumn{1}{|c|}{\multirow{7}[14]{*}{Malware}} & \textit{oldMal} & 2010-2012 & 2,000 & 1,833 & 1591 \\
\cline{2-6}    \multicolumn{1}{|c|}{} & \textit{Mal13} & 2013 & 415 & 399 & 335 \\
\cline{2-6}    \multicolumn{1}{|c|}{} & \textit{Mal14} & 2014 & 160 & 142 & 135 \\
\cline{2-6}    \multicolumn{1}{|c|}{} & \textit{Mal15} & 2015 & 119 & 115 & 115 \\
\cline{2-6}    \multicolumn{1}{|c|}{} & \textit{Mal16} & 2016 & 203 & 184 & 146 \\
\cline{2-6}    \multicolumn{1}{|c|}{} & \textit{Mal17} & 2017 & 104 & 102 & 101 \\
\cline{2-6}    \multicolumn{1}{|c|}{} & \textit{\color{blue}{total}} & \multicolumn{1}{r|}{\textit{\color{blue}{2010-2017}}} & \multicolumn{1}{r|}{\textit{\color{blue}{3,001}}} & \multicolumn{1}{r|}{\textit{\color{blue}{2,775}}} & \textit{\color{blue}{2,423}} \\
    \hline
    All apps & \textbf{total} & \multicolumn{1}{r|}{\textbf{2010-2017}} & \multicolumn{1}{r|}{\textbf{6,432}} & \multicolumn{1}{r|}{\textbf{5,934}} & \textbf{5,098} \\
    \hline
    \end{tabular}%
  \label{tab:datasets}%
\end{table}%

To understand how the sensitive accesses in Android apps evolve, we
collected real-world app samples from 
early 2010 to mid 2017, and characterized them in terms of their SAD profiles.
We first introduce the characterization datasets and then highlight the major
findings from the study.

\subsection{Datasets}
Table~\ref{tab:datasets} gives an overview of our study datasets, 
including numbers of all samples downloaded ({\em\#A.S.}), samples 
characterized ({\em\#C.S.}), and samples used in the evaluation of our technique ({\em \#E.S.}). 

The first set of 1,221 benign apps ({\em oldBen}) was collected by downloading the top 50 popular apps in each
app category
on Google Play in late 2014. 
The new benign data set ({\em newBen}) was obtained by downloading the top 100 popular apps in
each of the app categories on Google Play by May, 2017.
The {\em oldMal} malware set was obtained by randomly selecting 2,000 apps
from the Drebin dataset~\cite{Arp14}.
The {\em Mal17} set consists of 104 new malware in the wild we manually collected from various sources.
The malware sets for 2013 through 2016 consists of all apps from respective years in the
AndroZoo dataset~\cite{allix2016androzoo} that were identified as malware by $\ge32$ AV tools (out of 63) of
VirusTotal~\cite{virusTotal} as recorded there.

Our entire datasets include 3,431 benign and 3,001 malicious samples, for a total of 6,432 benchmarks.
Constructing the SAD profile for some of these benchmarks were not successful. Most of
those failing cases were due to missing assets in APKs and thus cannot be executed for tracing.
Among others, some either have corrupted APKs (which cannot be unpacked for analysis) or cannot be instrumented due to
the inability of Soot to process their bytecode; the rest cannot be installed or did not produce a
a useful trace (i.e., having non-zero line coverage).
The {\em\#C.S.} column of Table~\ref{tab:datasets} lists the numbers of samples in each dataset for which
we were able to obtain a meaningful SAD profile (we used one profile per benchmark app).
These apps ({\em valid samples}), a total of 5,934, were used in our characterization study. 
We elaborate on the numbers of the last column in Section~\ref{subsec:evaldata}.

\subsection{Experimental Settings}\label{subsec:studysetup}
We produced the SAD profile of each valid sample by running the instrumented app
on an Android emulator (Nexus One), with SDK 6.0 (API 23), 2G RAM, and 1G SD storage.
The emulator ran on an Ubuntu (15.04) desktop of 16G memory and 8-core 2.6GHz CPU.
Each app was exercised for ten minutes by the random inputs generated by Monkey~\cite{monkey}.
We used Monkey because it achieves an average code coverage at least comparable to other existing
alternatives (research prototypes) while
with the best usability and reliability.
We chose $T=10$ instead of a longer time in consideration of the time expense of the study given the large number of apps,
and the previous finding that Monkey attained its highest coverage in ten minutes in an average case~\cite{choudhary2015automated}.
To minimize bias, we run each app in a fresh emulator setting (by restarting a clean emulator for each app).

The total trace size of all valid sample SAD profiles is about 128GB.
Collecting the traces took over an effective time of three months.
The line coverage of the per-sample traces ranged from
1.8\% to 97.1\%, for a mean of 48.2\%, standard deviation of 22.1\%, and median of 50.1\%.
With our toolkit, a real Android device can be easily plugged in to produce traces
for {\tech} as well. We used an emulator in order to test the robustness of our dynamic
detection approach to malware evasion techniques targeting dynamic analysis (e.g., hiding
malicious behaviors when they detected being run on an emulator)~\cite{rasthofer2017making}.


\subsection{Evolutionary Characteristics}\label{sec:charresults}
We present the most important results of our characterization, based on the
SAD profiles of 5,934 apps (3,159 benign and 2,775 malicious) over the past seven years.
To understand the evolutionary patterns of sensitive accesses of Android apps, we
computed the summary statistics for each of the 8 datasets separately and compare
among them.

\subsubsection{Evolution in Use Extent}
\begin{figure*}
  \centering
  \begin{tabular}{cc}
    \includegraphics[scale=0.5]{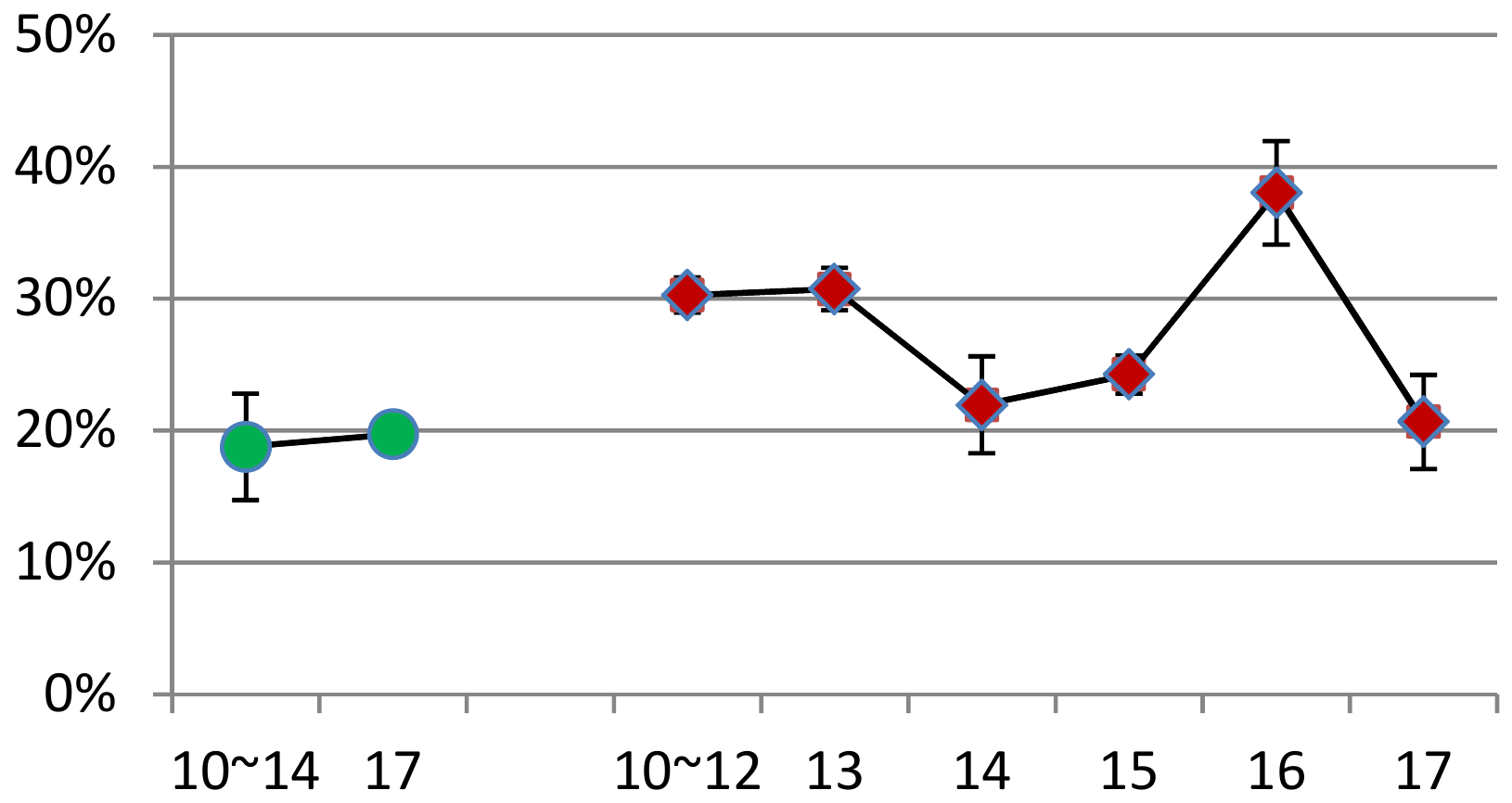} &
    \includegraphics[scale=0.5]{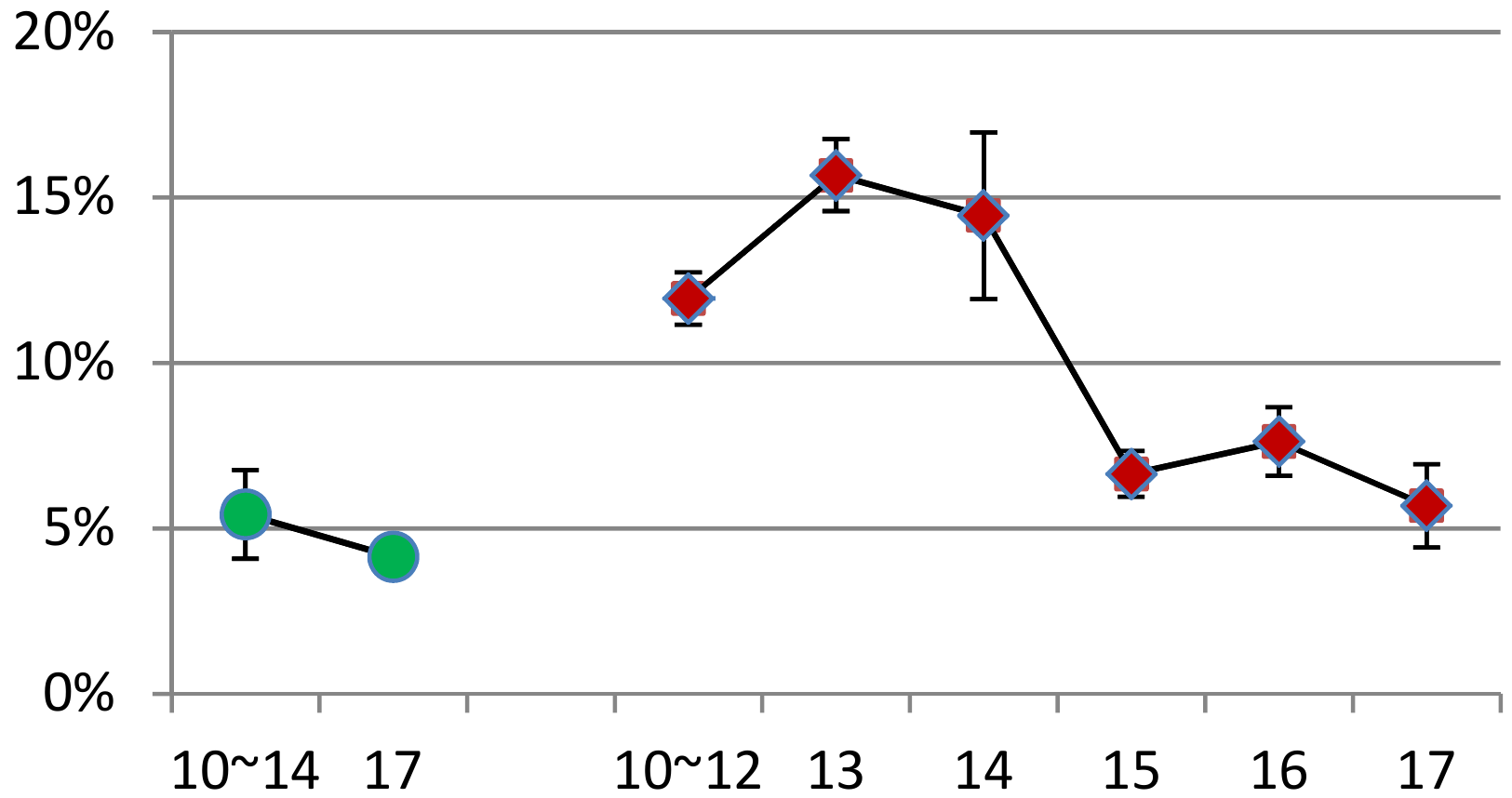} \\
    (a) Source callsite & (b) Sink callsite \\
    \includegraphics[scale=0.5]{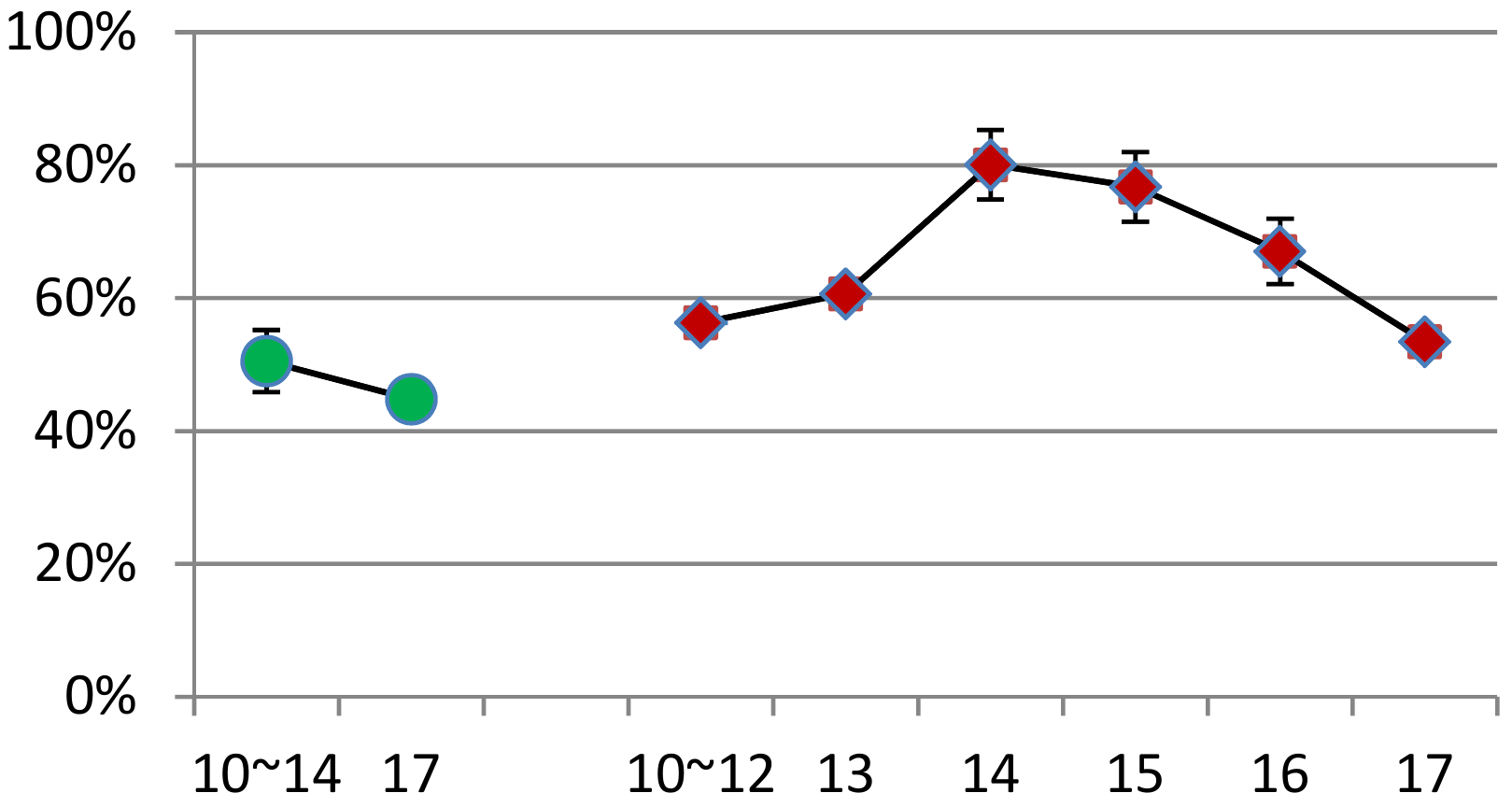} &
    \includegraphics[scale=0.5]{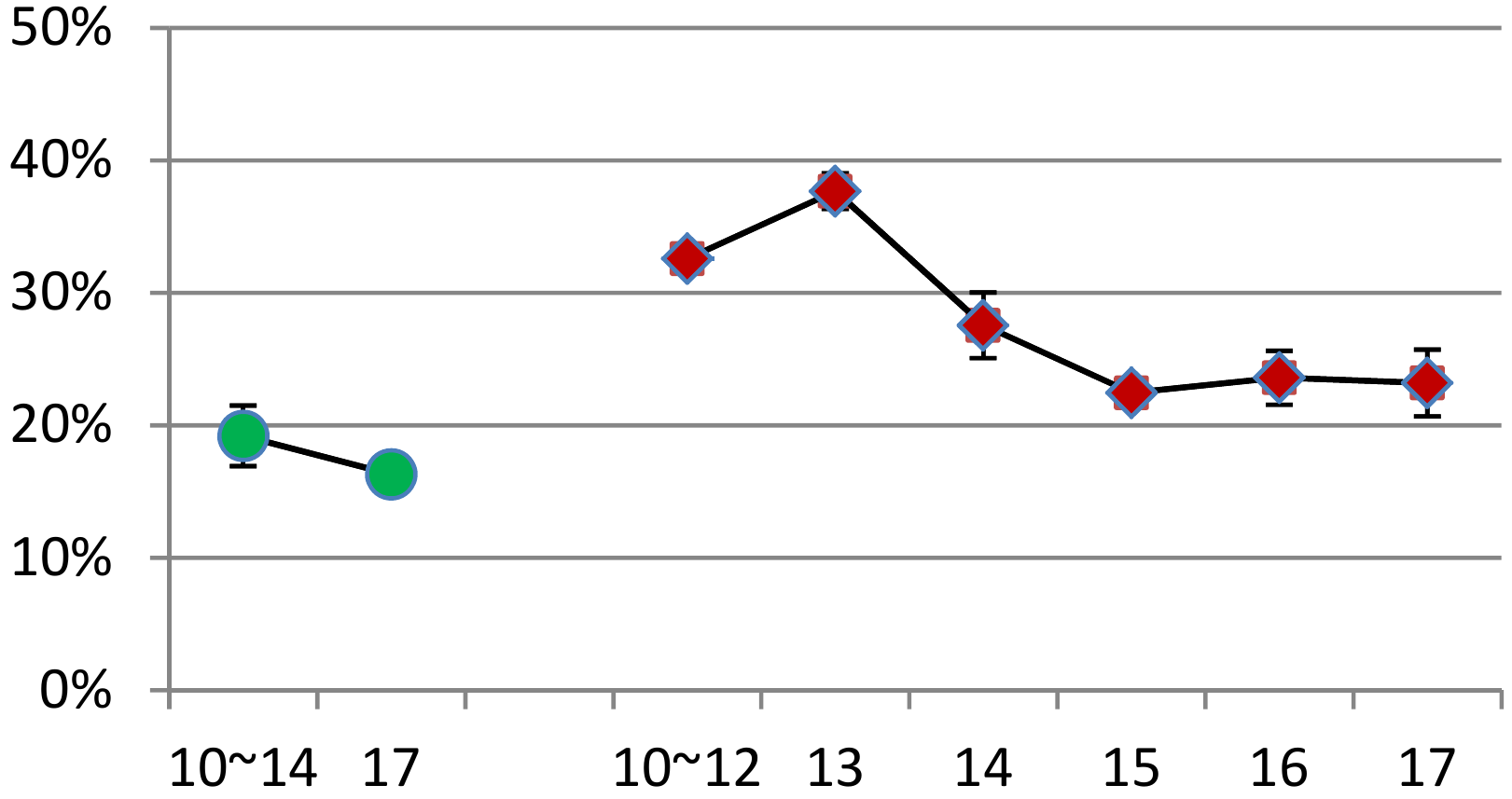} \\
    (c) Source call instance & (d) Sink call instance \\
    \includegraphics[scale=0.5]{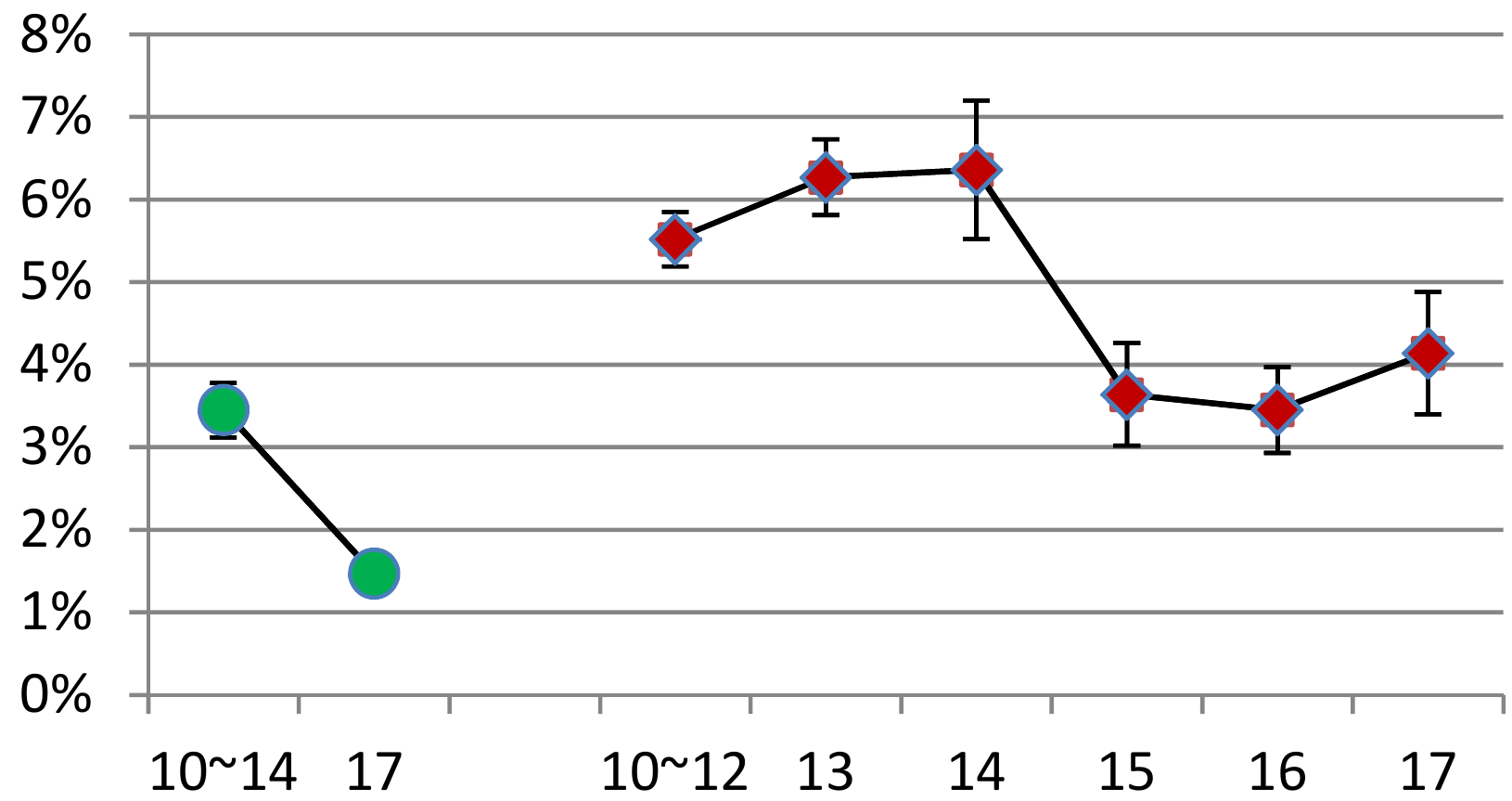} &
    \includegraphics[scale=0.5]{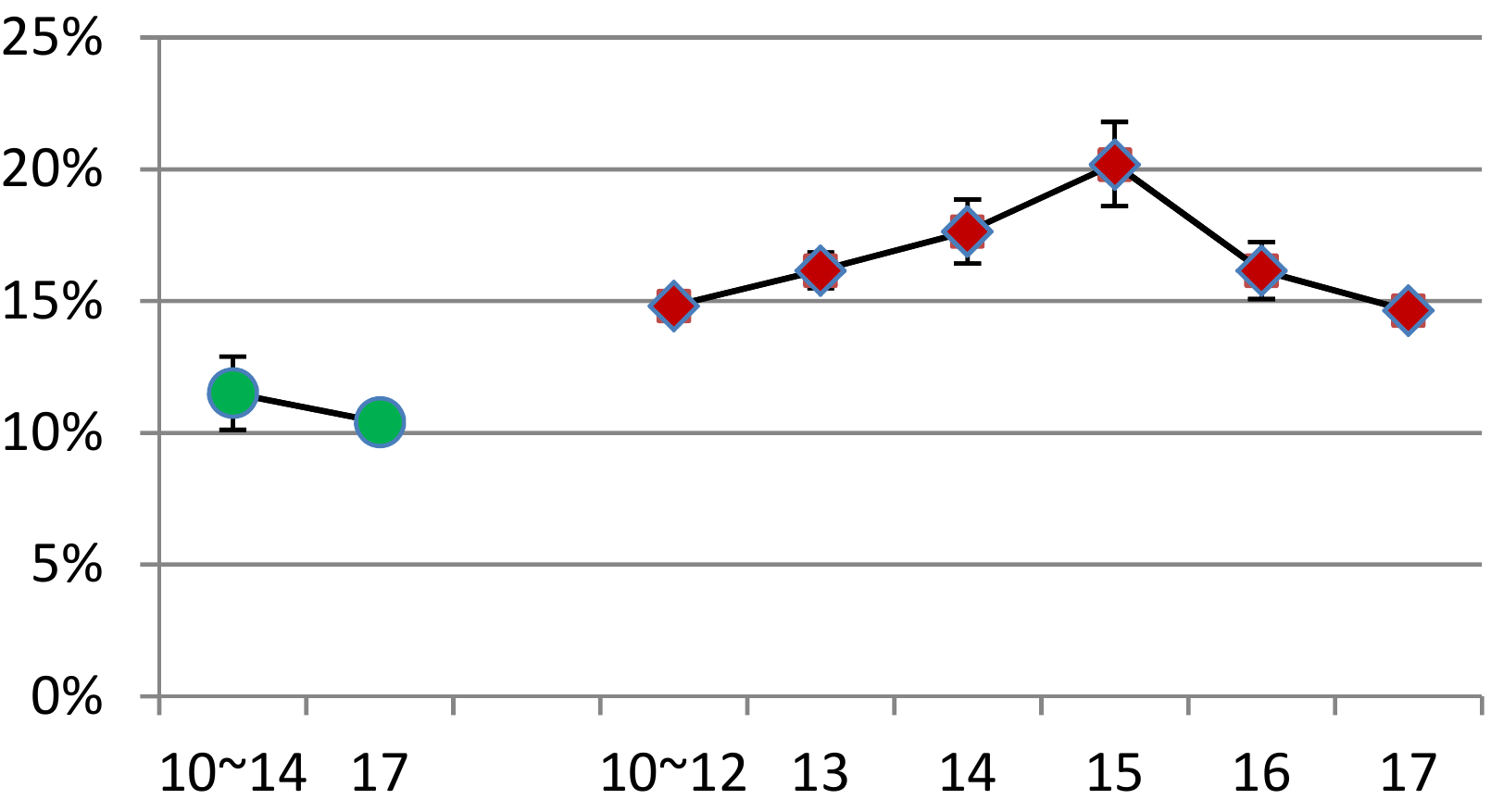} \\
    (e) Vulnerable source callsite & (f) Vulnerable sink callsite \\
    \includegraphics[scale=0.5]{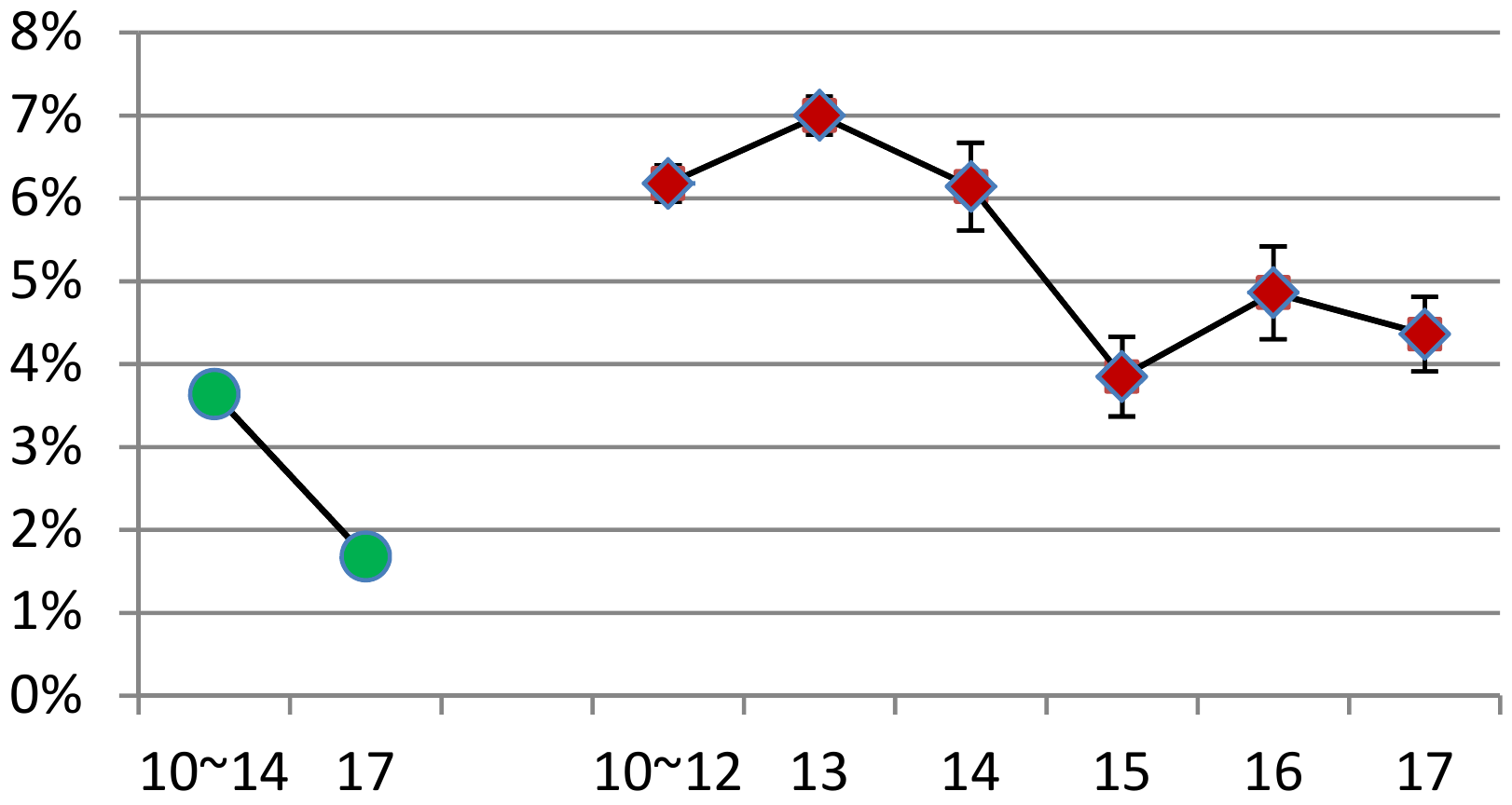} &
    \includegraphics[scale=0.5]{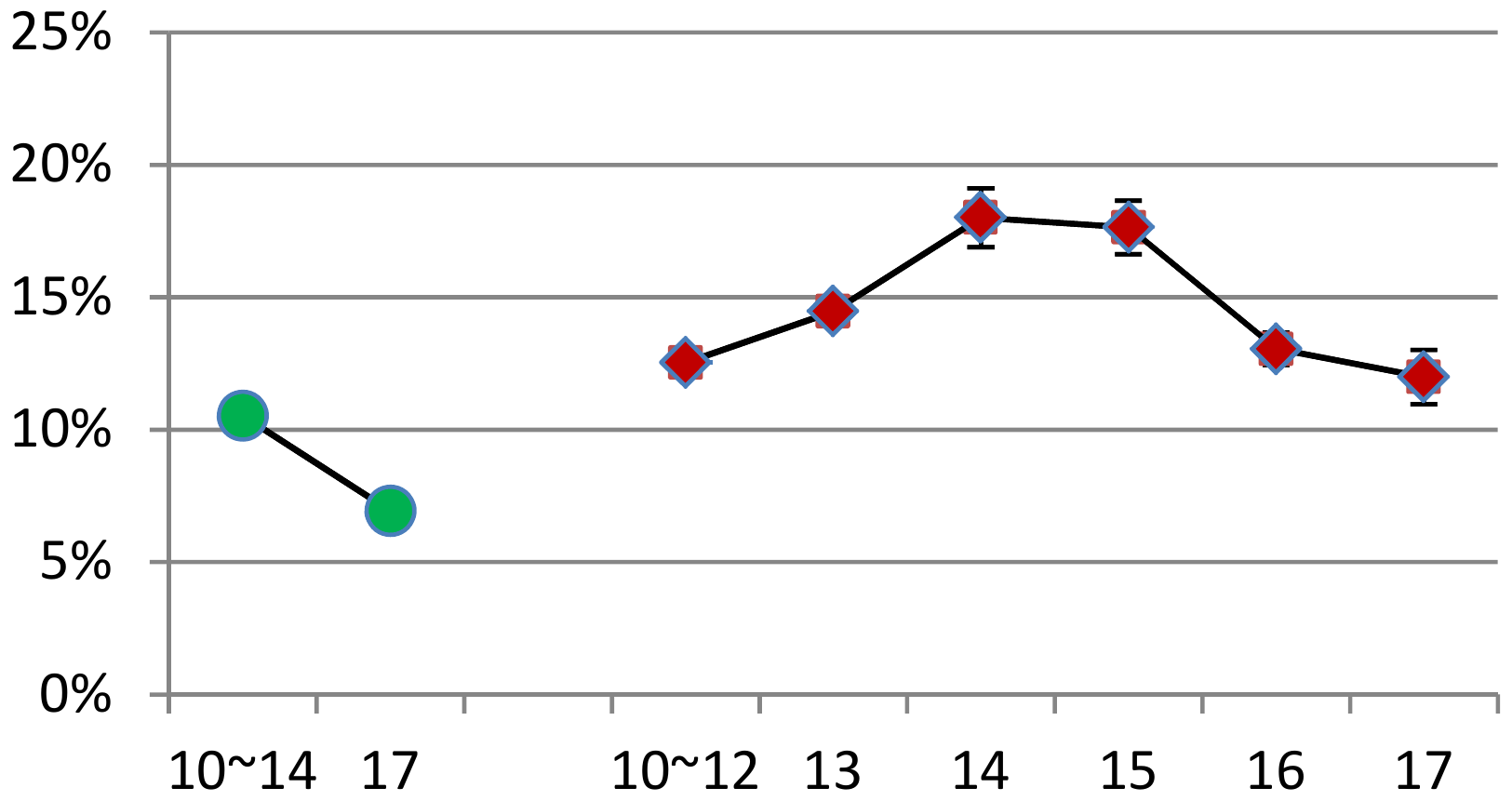} \\
    (g) Vulnerable source call instance & (h) Vulnerable sink call instance \\
  \end{tabular}
  \vspace{-4pt}
  \caption{Mean percentage of calls for sensitive accesses, $y$ axis) over years ($x$ axis) in benign apps (circle) and malware (diamond). Error bars show the 0.95 confidence intervals of the means.}
  \label{fig:srcsinkuse}
  \vspace{-4pt}
\end{figure*}

Figure~\ref{fig:srcsinkuse} shows the patterns regarding the extent of those accesses
in terms of source and sink calls in both the {\em callsite} and {\em instance} views.
In each chart, we show the average-case SAD profiles for benign versus malicious apps,
separately, with respect to all the characterization datasets. Since we are interested in
comparing the means, we computed the 0.95 confidence interval for each mean to estimate
how these sample means approximate those of the respective populations.
Data points underlying these means are values of corresponding features (see Table~\ref{tab:datasets}).

\setlength{\tabcolsep}{2pt}

The first four charts indicate that the sensitive accesses were substantial
(given our conservative definition of such accesses) in all the benchmarks studied, in either views.
Among all method invocations, calls to sources were even more extensive than calls to sinks (20-80\% versus 5-40\%),
partly because the predefined list of sources is much larger than the list of sinks we referred to.
Partly due to the same reason, as the next four charts show, percentages of sources that can reach sinks (i.e., vulnerable
sources) were much lesser compared to vulnerable sinks (1-7\% versus 10-20\%): the dominators for the former were much larger.
As expected, benign apps had generally less extensive accesses to sensitive information and operations than malware, both in
terms of the diversity of source/sink APIs invoked (i.e., the {\em callsite} view) and the frequency of those API calls
(i.e., the {\em instance} view).
In contrast to these numbers themselves, however, the more important is the evolutionary trends revealed by the comparison of
these numbers over the years, which we discuss below.

\vspace{4pt}\noindent\textbf{Evolution of benign apps.}
While we only studied two benign datasets, they were at least 3 years apart.
Figure~\ref{fig:srcsinkuse} shows that over this period of time, sensitive accesses in
benign apps tend to drop, but only slightly (by less than 5\%) in terms of callsites and call instances
targeting sinks and call instances targeting sources. An exception was the trend in terms of
source callsites (chart (a)): benign apps of 2017 had tiny increase in this feature compared to
three years ago. Thus, it appeared that benign apps were making a little more diverse source API calls,
but less frequently overall ((c)).

\vspace{4pt}\noindent\textbf{Evolution of malware.}
Malware experienced ups and downs in sensitive accesses with respect to any of the 8 metrics
depicted in Figure~\ref{fig:srcsinkuse}. Interestingly, over the years, malicious apps tend to
increase in the extent of these accesses at some point but then drop, eventually lower than the early time.
Consider vulnerable sink call instances (chart (h)) for example, from 2010 through 2014, the extent
grew continuously (from 12 to 18\%) and then kept decreasing afterwards (by 11\%).
Despite the different scales ($y$ axis), the gaps between 2010 and 2017 revealed that the malware reduced
significantly in the diversity of sensitive APIs invoked, but relatively lesser in the frequency of such calls.
An implication of this observation is that malware tended to concentrate on a smaller set of APIs for sensitive access.

\vspace{4pt}\noindent\textbf{Benign versus malware.}
Most importantly, regardless of the fluctuations seen by both malware and benign apps,
the differences in the extent of sensitive accesses between the two groups were very consistent:
a visual horizontal dividing line between them could be drawn in any of the charts of Figure~\ref{fig:srcsinkuse}.
In other words, while both have changed within the group, the patterns of the differences appeared to be quite stable,
suggesting the 8 metrics to be able to consistently differentiate malware from benign apps.

\subsubsection{Evolution in Dominating Categories}
\begin{figure*}
  \centering
  \begin{tabular}{cc}
    \includegraphics[scale=0.5]{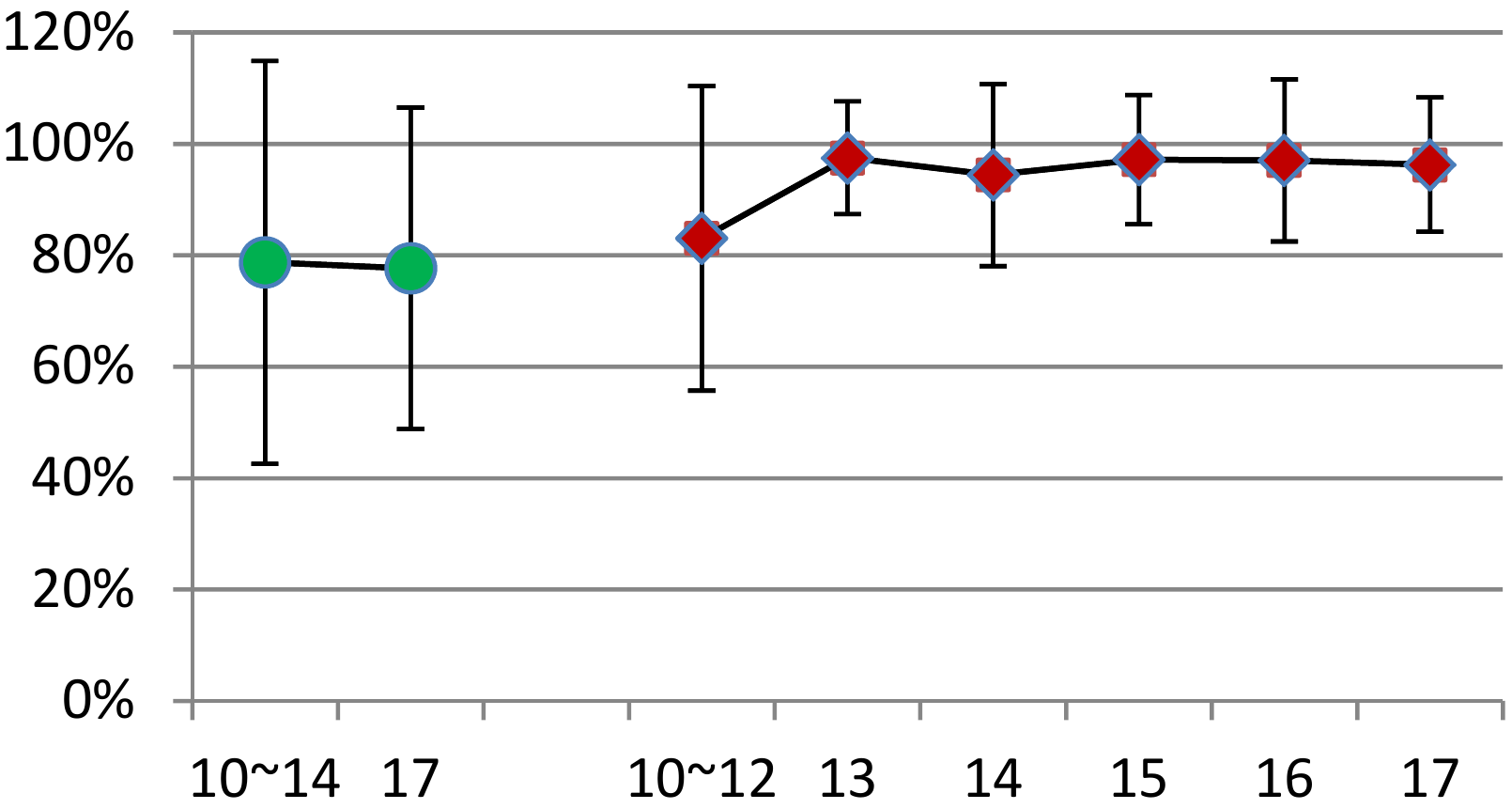} &
    \includegraphics[scale=0.5]{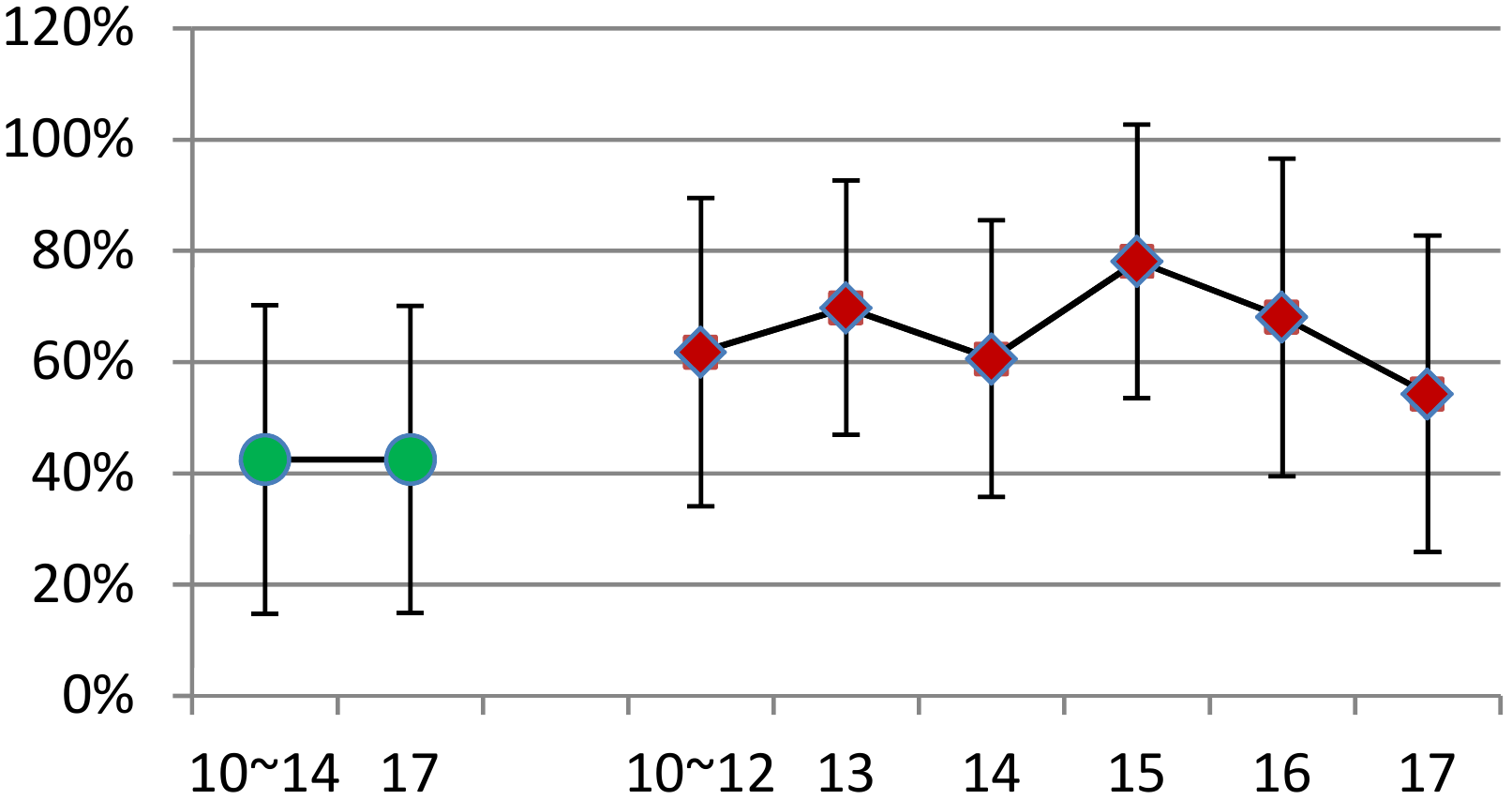} 
  \end{tabular}
  \vspace{-4pt}
  \caption{Mean percentage of source call instances ($y$ axis) that accesses network information (left), and of sink call instances ($y$ axis) that perform account settings (right), over years ($x$ axis) in benign apps (circle) and malware (diamond). Error bars show the 0.95 confidence intervals of the means.}
  \label{fig:srcsinkcat}
  \vspace{-4pt}
\end{figure*}

Most of the features in our SAD profile concern the categories of source/sink APIs associated with
the sensitive accesses. Our results clearly indicate that in both malware and benign apps, among
all sources exercised, the dominating ones were those retrieving network information, while calls to sinks
related to account settings dominated all sink calls. This observation was consistent regardless of the views and
whether the sources/sinks were vulnerable or not.

Figure~\ref{fig:srcsinkcat}, in the format as Figure~\ref{fig:srcsinkuse}, depicts the evolutionary trends concerning these two dominating categories of sensitive access, in terms of all call instances to sources (left) and sinks (right).
Compared to the use extent, the means had wider confidence intervals, meaning that these numbers were more approximate and there were
larger variations among individual apps within each dataset.
Nevertheless, the average-case metrics still clearly reveal consistent patterns as seen above.
Sensitive accesses in benign apps exhibited an almost unchanged focus on the two top categories.
In contrast, malware again saw more variations over the years, experiencing ups and downs like in the overall use extent.
However, the fluctuations were much smaller in general.
Finally and most importantly, the clear division between the benign apps and malware was quite consistent over years (albeit
the gaps were larger at some points than at others).
We have observed similar patterns in benign apps and malware separately, and in the consistent differentiation between
the two groups, in other source/sink categories as well.

\section{The {\tech} System}
\label{sec:tech}
This section presents {\tech}, our novel Android app classification approach using 
supervised learning based on the defining features of the SAD profile. 

\vspace{4pt}\noindent\textbf{Overview of approach.}
{\tech} is a dynamic malware detection system based on the 
SAD profile. Like existing ML-based app classification systems, {\tech} works in 
two major phases: training and testing. 
In the training phase, it relies on a set of labeled app samples 
to build a prediction model. Then in the testing phase, the trained model 
is applied to classify novel apps. Our main goal with and motivation for {\tech}, 
however, is that this model can be used effectively (i.e., with high classification performance) 
some time after it is trained, so that {\em emerging} malware can be discovered by the system without 
being retrained on samples for the new malware.  
Intuitively, the longer the time span the better, and more sustainable the system. 

To make the prediction model sustainable, the key is that the underlying features 
can, for a relatively long period of time, distinguish benign apps from malware 
To that end, {\tech} uses the features that define the SAD profile of an Android app (Table~\ref{tab:features}). 
Our longitudinal characterization (Section~\ref{sec:charresults}) shows that, with 
the evolution of malware attack strategies and the Android system, benign apps and malware developed in different years 
tends to exhibit generally similar behaviors in terms of their SAD profile. Moreover, the discrimination between both groups 
appeared to be clearly consistent. 
These findings provide an essential grounding for {\tech} to achieve 
sustainable classification over years. 

\vspace{4pt}\noindent\textbf{Feature extraction.}
Extracting the SAD features is needed both for obtaining the sample set for training {\tech} 
and for later predicting the label of a novel app using {\tech}. 
We compute the features of all training and testing apps as described in Section~\ref{sec:bkg}.

\vspace{4pt}\noindent\textbf{Classification.}
To find the best ML model for {\tech}, we have performed a comparative study 
of {\tech} variants using most of the supervised classification models provided by the
Scikit-learn library~\cite{pedregosa2011scikit}. These models include: 
random forest (RF), support vector machines (SVM) with both linear and radial basis function kernels, decision trees/C4.5 (DT), $k$-nearest neighbours ($k$NN), naive Bayes (NB) with three possible probability models 
(Gaussian, Multinomial Bernoulli), AdaBoost, Gradient Tree Boosting, Extra Trees, and the Bagging classifier. 
By comparing these variants of {\tech} with respect to classification performance (precision, recall, and F1-measure), 
we found that RF consistently achieved the best performance on any of our datasets. 
Thus, we chose RF as the learning algorithm for {\tech}. 
%
{\tech} is first trained on benign apps labeled as BENIGN 
and malicious apps all labeled as MALICIOUS, and then classifies a novel app as either of the two labels. 
\section{Evaluation}
We assessed {\tech} in terms of classification performance and efficiency, 
versus {\base}. We chose {\base} as the baseline approach because it 
is a state-of-the-art ML-based malware detection system 
and shares a similar goal to ours (i.e., sustainable detection). 
Also, both systems use features based on API calls in an app, albeit {\base} 
extracts features with static analysis, using FlowDroid~\cite{Arzt14} to build 
call graphs while {\tech} is based on pure dynamic analysis. 

Our evaluation consists of two main separate studies. In {\em Study I}, we 
assess the performance of {\tech} with training and testing apps developed in a 
same period of time (referred to as {\em same-period detection}). 
In {\em Study II}, we focus on evaluating the sustainability 
of {\tech} by assessing its performance when it is trained on older datasets and 
tests newer ones, spanning one to five years (referred to as {\em detection over time}). 
%
%
We performed an additional study, {\em Study III}, to assess the resiliency of {\tech} to 
sophisticated obfuscation schemes adopted in modern Android apps (malware in particular).

\subsection{Used Benchmarks}\label{subsec:evaldata}
We started with the same valid samples as described in (the {\em\#C.S.} column of) Table~\ref{tab:datasets}. 
{\base} failed to analyze some samples due to the limitation of its underlying analysis by 
FlowDroid (as the authors discussed~\cite{mariconti2017mamadroid}). 
For a fair comparison, we removed those samples 
to make sure the comparison is based on exactly the same datasets.
The {\em\#E.S.} column of Table~\ref{tab:datasets}  
lists the remaining samples in each dataset that were eventually  
used in our evaluation studies. 

Assessing the performance of both techniques 
needs mixed datasets each consisting of both malware and benign apps. 
Thus, we used several combinations of characterization datasets for our evaluation. 
Specifically, for {\em Study I}, 
we combined the {\em oldBen} set with relatively older malware sets ({\em oldMal}, 
{\em Mal13}, and {\em Mal14}), and {\em newBen} set with relatively newer malware sets 
({\em Mal15}, {\em Mal16}, and {\em Mal17}). Thus, we had {\em six mixed datasets} each covering 
a different malware set. In addition, we consider all the available benchmarks constituting 
an {\em all-data} set. 

For {\em Study II}, we used the combinations of malware and benign datsets such that we 
take the oldest malware set ({\em oldMal}) for training to test each of the 
newer malware sets. In this way, we can test the sustainability of the two techniques 
for the longest possible time span with respect to our datasets. 
We used the two benign sets, separately for two repetitions of this study.

We used an additional dataset, named {\em MalObf}, for {\em Study III}. This dataset 
was obtained from the Contagio subset in the Praguard obfuscation benchmark suite~\cite{maiorca2015stealth}. 
It initially includes 237 malicious apps that each adopted three obfuscation schemes together: 
encryption, renaming, and reflection. 
For 17 of these apps, {\tech} failed to produce valid SAD profiles. Thus, {\em MalObf} consists of 
220 samples used in {\em Study III}. 

\subsection{Methodology}
In our evaluation, we gauged the performance of {\tech} versus the baseline in terms of three 
metrics: precision, recall, and F1-measure (accuracy), as defined below. 

Give an app $p$ under classification,
$p$ is a {\em true positive} (TP) if the true label is $L$ (i.e., BENIGN or MALICIOUS) and the predicted class label is $L$;
it is a {\em false positive} (FP) if the true label is not $L$ and the predicted label is $L$;
it is a {\em true negative} (TN) if the true label is not $L$ but the predicted label is not $L$,
it is a {\em false negative} (FN) if the true label is $L$ and the predicated label is not $L$.
Accordingly, 
regardless of the number of classes (different labels) in the model:
{\small
Precision ($P$) = $\frac{TP}{TP+FP}$,
Recall ($R$) = $\frac{TP}{TP+FN}$}, 
and {\small F1 = $2*\frac{P*R}{P+R}$}. 
For evaluating malware detection, these metrics are computed concerning $L$=MALICIOUS only.
%


In {\em Study I}, for each of totally seven evaluation datasets, we 
compute results from a 10-fold CV to facilitate comparison with the baseline (which
was partially evaluated so originally). 
We further performed an independent testing using the biggest dataset ({\em all-data}). 
Specifically, we randomly selected 30\% samples from each class (malware or benign) in this dataset 
and reserved them as unseen/novel samples. 
By doing so, we intend to test the capability of both techniques in classifying 
new apps (albeit from the same time period as the training set) and to complement the CV results. 

For {\em Study II}, we used each of the mixed evaluation datasets considered for this study,  
as either training or testing data. Since there were no intersections between each pair of these training and testing datasets, 
this study essentially performs independent testings.

In {\em Study III}, we used the entire {\em MalObf} set for testing and the {\em oldBen+oldMal} set for training. 
These two datasets do not have any common apps. Thus, this study also performs an independent testing. 

Both techniques used the same RF algorithm. For {\tech}, we used 100 estimators and left other parameters 
as default in Scikit-learn. For {\base}, we used the parameters as described in~\cite{mariconti2017mamadroid}. 
MamaDroid was developed to work in two modes. For ease of comparison, we simply report the best results of any mode in our 
evaluation results for MamaDroid. 
We used the same experimental settings for all evaluation studies as for 
the characterization (Section~\ref{subsec:studysetup}).


\subsection{Study I: Same-Period Detection}
In this study, we aimed to gauge the performance for
classifying apps developed in the same period of time as were the training samples.
As mentioned earlier, we used both 10-fold CV and independent testing for this evaluation.

%
%
\setlength{\tabcolsep}{1.5pt}
\begin{table}[tp]
  \centering
  \caption{Comparison of cross-validation results between the two techniques.}
    \begin{tabular}{|r|r|r|r||r|r|r|}
    \hline
    \multicolumn{1}{|c|}{\multirow{2}[4]{*}{Dataset}} & \multicolumn{3}{c||}{\textit{{\tech} (our work)}} & \multicolumn{3}{c|}{\textit{{\base}~\cite{mariconti2017mamadroid}}} \\
\cline{2-7}    \multicolumn{1}{|c|}{} & P & R & F1  & P & R & F1 \\
    \hline
    all-data & 0.937 & 0.937 & 0.937 & \textbf{0.945} & \textbf{0.942} & \textbf{0.943} \\
    \hline
    oldBen+oldMal & 0.948 & 0.948 & 0.946 & \textbf{0.960} & \textbf{0.960} & \textbf{0.958} \\
    \hline
    oldBen+Mal13 & \textbf{0.948} & \textbf{0.948} & \textbf{0.946} & 0.926 & 0.926 & 0.923 \\
    \hline
    oldBen+Mal14 & \textbf{0.939} & \textbf{0.933} & \textbf{0.929} & 0.935 & 0.931 & 0.923 \\
    \hline
    newBen+Mal15 & 0.981 & 0.981 & 0.980 & \textbf{0.982} & 0.981 & 0.980 \\
    \hline
    newBen+Mal16 & \textbf{0.986} & \textbf{0.986} & \textbf{0.985} & 0.970 & 0.972 & 0.970 \\
    \hline
    newBen+Mal17 & \textbf{0.977} & \textbf{0.980} & \textbf{0.974} & 0.971 & 0.972 & 0.968 \\
    \hline
    \hline
    \multicolumn{1}{r|}{Overall average} & \textbf{0.959} & \textbf{0.959} & \multicolumn{1}{r|}{\textbf{0.956}} & 0.956 & 0.955 & \multicolumn{1}{r}{0.952} \\
    \end{tabular}%
  \label{tab:cvresults}%
\end{table}%

\vspace{4pt}\noindent\textbf{Cross validation.}
Table~\ref{tab:cvresults} lists the validation results of our approach versus the baseline, in terms of
the three classification performance metrics described earlier: precision ({\em P}),
recall ({\em P}), and F1 measure ({\em F1}). Each row of the table shows these metrics for both techniques obtained from
one of the seven datasets (first column), except for that the last row shows the simple averages of the numbers
in the same column. To facilitate comparison, on any dataset, the higher metrics are marked in bold face.
Note that each number was the average of numbers of the same metric across the ten passes of the 10-fold CV, and the
F1 measures were the averages of per-pass F1 instead of being computed from the averaged precisions and recalls.

Overall, the numbers between the two compared approaches are very close for any specific dataset and individual metrics.
In general, both techniques performed well for same-period detection, achieving 93~99\% F1 accuracy.
Either technique had better performance at some datasets than the other.
By overall average, {\tech} had slightly higher performance than the baseline.
However, the differences were quite small, especially in terms of the F1 measure.
In sum, the cross validation suggests that both techniques have no significant difference in classification
performance.

\vspace{4pt}\noindent\textbf{Independent testing.}
As a known limitation, cross validation may suffer from biases in its results due to possible overfitting.
To complement with an independent testing,
we used 70\% of benchmarks in the {\em all-data} set to train both techniques, and then tested against
them with the other samples in this set.
Our results show again similar results between the two techniques compared, with {\tech} obtaining a 95.6\%
F1 accuracy versus the baseline number of 94.7\% by {\base}.
Thus, the independent testing confirmed that both techniques performed almost equally well (without
significant difference).

In all, through two evaluation strategies, we note that
{\tech} can obtain high-accuracy for same-period detection, comparable to
a state-of-the-art approach {\base}. 
It is also worth mentioning that, despite the different datasets used, 
the same-period detection performance results we obtained in 
this study were almost the same to what were reported in the original evaluation of 
{\base}~\cite{mariconti2017mamadroid} (93-99\% versus 92-98\% of F1), 
with respect to the same validation method (i.e., 10-fold CV).


\setlength{\tabcolsep}{1pt}
\begin{table*}[tp]
  \centering
  \caption{Comparison of performance between the two techniques for detection over time}\vspace{-8pt}
    \begin{tabular}{|r|r|r|r|r|r|r|r|r|r|r|r|r|r|r|r|r|}
    \hline
    \multicolumn{1}{|c|}{\multirow{3}[6]{*}{\shortstack{Training\\dataset}}} & \multicolumn{1}{c|}{\multirow{3}[6]{*}{Techniques}} & \multicolumn{15}{c|}{Testing dataset} \\
\cline{3-17}    \multicolumn{1}{|c|}{} & \multicolumn{1}{c|}{} & \multicolumn{3}{c|}{oldBen+Mal13} & \multicolumn{3}{c|}{oldBen+Mal14} & \multicolumn{3}{c|}{oldBen+Mal15} & \multicolumn{3}{c|}{oldBen+Mal16} & \multicolumn{3}{c|}{oldBen+Mal17} \\
\cline{3-17}    \multicolumn{1}{|c|}{} & \multicolumn{1}{c|}{} & \multicolumn{1}{c|}{P} & \multicolumn{1}{c|}{R} & \multicolumn{1}{c|}{F1} & \multicolumn{1}{c|}{P} & \multicolumn{1}{c|}{R} & \multicolumn{1}{c|}{F1} & \multicolumn{1}{c|}{P} & \multicolumn{1}{c|}{R} & \multicolumn{1}{c|}{F1} & \multicolumn{1}{c|}{P} & \multicolumn{1}{c|}{R} & \multicolumn{1}{c|}{F1} & \multicolumn{1}{c|}{P} & \multicolumn{1}{c|}{R} & \multicolumn{1}{c|}{F1} \\
    \hline
    \multicolumn{1}{|c|}{\multirow{2}[4]{*}{\shortstack{oldBen+\\oldMal}}} & {\tech} & \textbf{0.999} & \textbf{0.979} & \textbf{0.989} & \textbf{0.999} & \textbf{0.881} & \textbf{0.937} & \textbf{0.999} & \textbf{0.875} & \textbf{0.933} & \textbf{0.999} & \textbf{0.870} & \textbf{0.930} & \textbf{0.999} & \textbf{0.709} & \textbf{0.830} \\
\cline{2-17}    \multicolumn{1}{|c|}{} & {\base} & 0.942 & 0.935 & 0.935 & 0.886 & 0.877 & 0.880 & 0.626 & 0.780 & 0.694 & 0.606 & {0.833} & 0.702 & 0.469 & 0.266 & 0.340  \\
    \hline
    \multicolumn{1}{|r}{} &   & \multicolumn{3}{c|}{newBen+Mal13} & \multicolumn{3}{c|}{newBen+Mal14} & \multicolumn{3}{c|}{newBen+Mal15} & \multicolumn{3}{c|}{newBen+Mal16} & \multicolumn{3}{c|}{newBen+Mal17} \\
    \hline
    \multicolumn{1}{|c|}{\multirow{2}[4]{*}{\shortstack{newBen+\\oldMal}}} & {\tech} & \textbf{0.997} & 0.973 & \textbf{0.985} & \textbf{0.991} & 0.837 & \textbf{0.908} & \textbf{0.984} & 0.750 & \textbf{0.857} & \multicolumn{1}{r|}{\textbf{0.978}} & \multicolumn{1}{r|}{0.800} & \textbf{0.880} & \textbf{0.971} & \textbf{0.698} & \textbf{0.812} \\
\cline{2-17}    \multicolumn{1}{|c|}{} & {\base} & 0.863 & \textbf{0.999} & 0.927 & 0.667 & \textbf{0.846} & 0.746 & 0.564 & \textbf{0.780} & 0.655 & 0.545 & \textbf{0.885} & 0.675 & 0.473 & 0.261 & 0.337 \\
    \hline
    \end{tabular}%
    \vspace{-8pt}
  \label{tab:overtimeresults}%
\end{table*}%

\subsection{Study II: detection over time}
Recall that the main objective of our technique is that 
it can sustain its detection capabilities (in terms of performance numbers) 
over time. 
Thus, this study aims at testing {\tech} against this objective by training 
it with old malware set and 
testing malicious apps developed in the following one to five years. 
We intended to see whether how using older versus newer benign samples 
in the training would affect the performance of sustainable malware detectors. 
Thus, we used both the old and new benign sets during the training, separately. 
Next, we examine the classification performance in this setting by looking at the numbers, 
and the sustainability, by focusing on the change trend in the performance over years.

\vspace{4pt}\noindent\textbf{Classification performance.}
Table~\ref{tab:overtimeresults} shows the detailed results on the classification performance 
of {\tech} against the baseline approach. 
Given each of the five relatively newer malware sets (from 2013 through 2017) used for testing and  
each of the two benign datasets used for training (along with the old malware set consistently), we had 
ten rounds of independent testing. In the table, the three metrics, precision ({\em P}), recall {\em R}, and 
F1 accuracy ({\em F1}) are listed for each round, and each of the two techniques compared. 
For example, when trained on {\em OldBen+oldMal} and tested on {\em oldBen+Mal16}, 
the precision, recall, and F1 of {\base} were 46.9\%, 26.6\%, and 34\%, respectively. 
For each round, the better performance numbers (between our work and the baseline) in any metric were highlighted 
by bold face.

The upper half of the table shows the results when using the old benign set for training. 
Both {\tech} and {\base} exhibited great 
performance (98\% versus 93\% F1) when detecting malware developed just one year 
after the trained malware samples were developed. 
For a time span of two years, the performance of both techniques started dropping by about 5\% in F1. 
Both techniques saw the decrease in both precision and recall. 
However, the reason was mainly because of the degraded precision (by 6\%) for {\base}, while for {\tech} the main 
cause was the decrease in recall (by 9\%). 
When detecting malware further into the feature, the results reveal that {\tech} remained 
highly accurately (with 93\% F1) up to four years. Even for a five-year span, it still attained a reasonable 
F1 accuracy of 83\%. Note that the precision was always high at 99\%, and it was the drop in recall that led to the 
the reduction in F1 accuracy. This is likely to be in line with the fact that dynamic analysis tends to give precise results 
with imperfect recall (since the analysis considers a subset of possible executions only). 
By contrast, {\base} did not sustain beyond the two year span, with F1 dropped to 70\% in four years and 34\% by the 
fifth year. The main causes continued to the decrease in precision, although for the five-year span 
it was the very low recall (26\%) that caused the poor F1 accuracy.  

The bottom half of the table shows the results for using the new benign sets in the training. 
In terms of the percentage numbers, the performance of both techniques was lower (by 2-5\% in F1) than when the 
old benign set was involved in training. Yet the comparative trends between the two techniques for each time span did not 
change much, nor did other observations from the upper half of the table (e.g., {\tech} mainly dropped in recall while {\base} 
in precision, except for the longest span). 
Overall, for detecting malware two years in the future, {\tech} attained 91\% F1 accuracy (versus 75\% by {\base}). 
The contrasts between our work and the baseline 
were 88\% versus 68\%, and 81\% versus 33\%, for four- and five-year spans, respectively. 

It is worth noting that we confirmed, using different datasets, 
that {\base} kept high performance (over 90\%) for one year only and 
reasonable performance (above 80\%) for at most two years: these were the findings from the original evaluation of MamaDroid.
 
\begin{figure}[tp]
  \centering
    \includegraphics[scale=0.35]{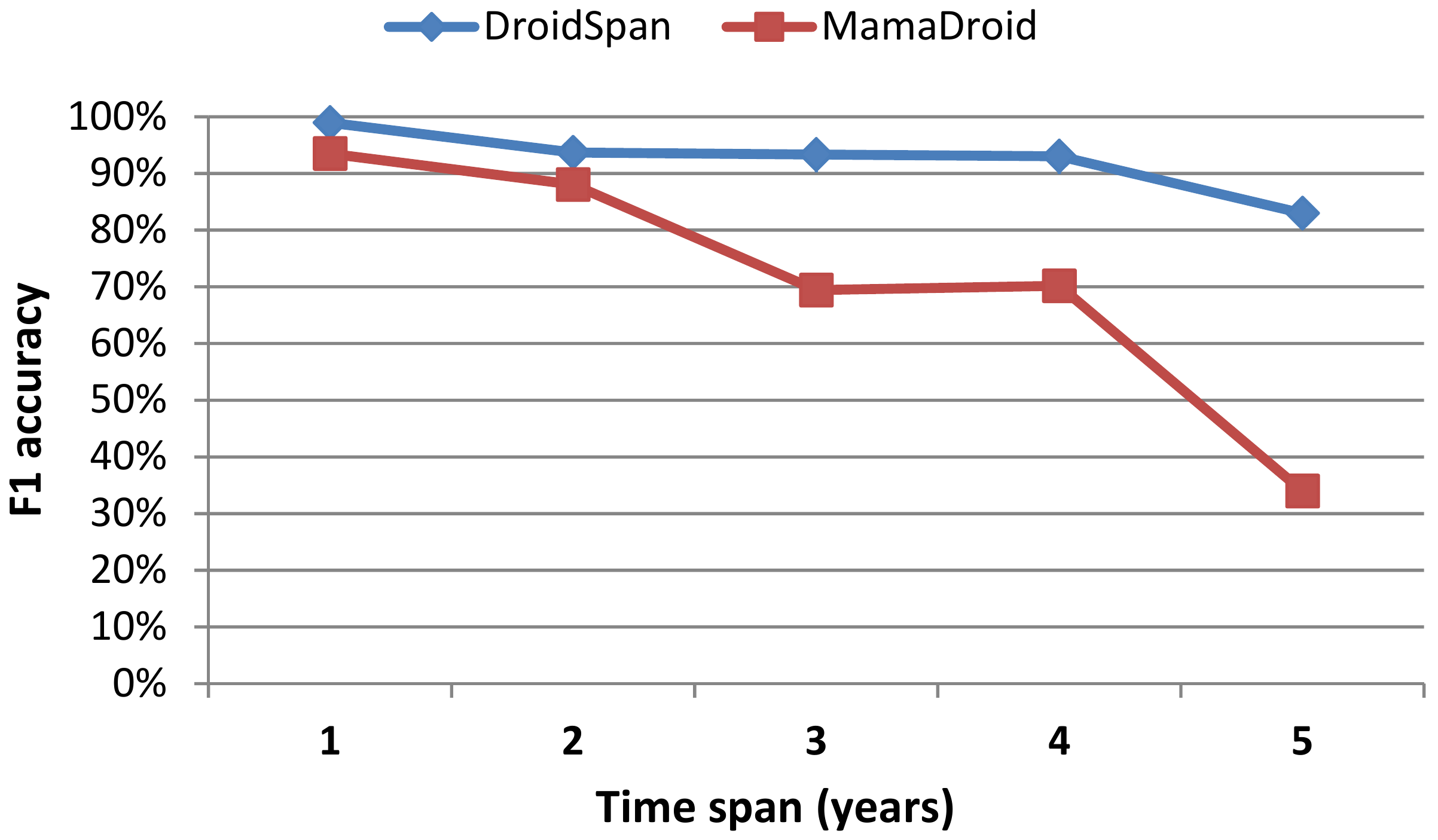} 
  \vspace{-15pt}
  \caption{Classification performance ($y$ axis) of {\tech} versus {\base}, trained on {\em oldBen+oldMal}, for detection over five years ($x$ axis).}
  \label{fig:spanoldben}
  \vspace{-10pt}
\end{figure}

\begin{figure}[tp]
  \centering
    \includegraphics[scale=0.5]{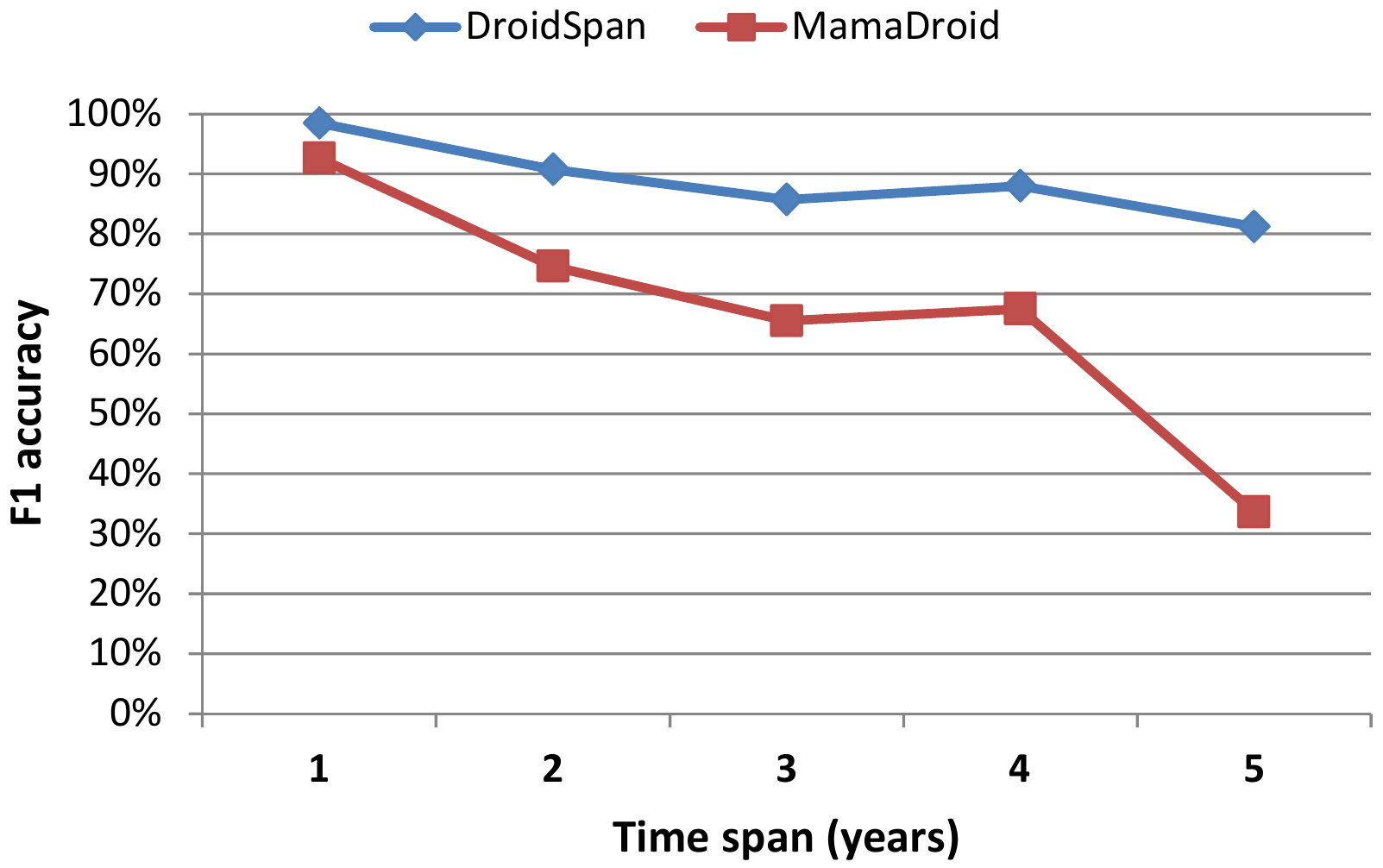}
  \vspace{-12pt}
  \caption{Classification performance ($y$ axis) of {\tech} versus {\base}, trained on {\em newBen+oldMal}, for detection over five years ($x$ axis).}
  \label{fig:spannewben}
  \vspace{-8pt}
\end{figure}

\vspace{4pt}\noindent\textbf{Sustainability.}
Figures~\ref{fig:spanoldben} and~\ref{fig:spannewben} depicts the trends of classification performance changes in terms of F1 accuracy with {\tech} versus {\base}, for using the old and new benign sets for training, respectively.
The plotted data points were the same F1 numbers from Table~\ref{tab:overtimeresults}. 
The two graphics clearly illustrate the overall sustainability patterns of both techniques in one to five years. As shown, 
while both classification approaches deteriorate over time, ours did so much slower than the baseline. 
Not only were the performance of {\tech} always higher than {\base} at any of the five time spans studied, overall 
the gap enlarged continuously with the increase in the length of the time span. 
These contrasts suggest that our work can sustain with high accuracy much longer than, thus 
clearly outperforms in sustainability, this state-of-the-art solution.


\subsection{Study III: Resiliency} 
Since many 
Android apps, malicious ones in particular, were obfuscated more 
or less (mainly in their code)~\cite{tam2017evolution}, it is crucial for a malware detector 
to be resilient to code obfuscation to be practically useful. 
As we discussed earlier, {\tech} is dynamic while monitoring all calls (including reflective 
and exception-driven ones) at runtime, it should be robust to typical code obfuscation schemes 
such as string encryption, reflection, and class/method renaming~\cite{maiorca2015stealth}.
To verify this hypothesis, in this study, we attempted to assess the resiliency of our technique 
to obfuscation against the baseline, using the {\em MalObf} dataset. 
All the three mentioned types of obfuscation schemes were applied in each app of this dataset. 

Trained on the {\em oldBen+oldMal}, {\tech} has been able to identify the malware in {\em MalObf}, with 
almost perfect performance (99\% precision, recall, and F1). This empirical result supports our 
hypothesis above. 
We also intended to do this same study on {\base}. 
However, the sophisticated analysis of FlowDroid~\cite{Arzt14} underlying the feature computation step 
of {\base} failed at any of the obfuscated apps due to the limitations of FlowDroid. 
While we cannot compare the performance between the two techniques against 
these obfuscated apps, as a static approach {\base} is likely to be more vulnerable to obfuscation than 
our dynamic technique.

\subsection{Efficiency Results}
We report the efficiency numbers with respect to all our datasets. 
The overhead of {\tech} is dominated by the cost for 
computing the SAD profiles of the training samples. 
The instrumentation took from 2 to 315 seconds 
(mean 33.9, std 36.7) per malicious app, with slightly (10s on average)
higher costs for benign apps. 
The profiling took a constant time of 10 minutes per app. 
Extracting SAD features from a trace took an average of 
15s for an app. 
Given the small feature set it used, {\tech} only spent 
very short time (within 3s) on model training and testing 
with the RF algorithm. 
Beyond the 128G traces in total for all the 5,934 apps (we currently 
simply stored all traces in plain text), other storage costs 
were trivial (i.e., 7M for storing all the feature files). 

On the same datasets, we experienced efficiency difficulties with {\base} in all 
its three steps. The static analysis cost of FlowDroid is known to be substantial. 
Despite a configuration for low analysis precision it used, {\base} took about 16 days
to finish the call-graph construction for our benchmarks (mean of 4min per app). 
For its best-performing (i.e., {\em package}) mode, {\base} 
also appeared slow with feature computation due to its 
very-high memory consumption as discussed in~\cite{mariconti2017mamadroid} (recall it has 150K feature values per app). This step 
took almost 3 days in our experiments, resulting in 11G feature files. 
The last step, training and testing, again was expensive because of loading the large 
feature files. For example, it spent about an hour for training on the {\em newBen+oldMal}
mixed dataset. These numbers, even considering the per-app averages, 
are much higher than those reported in~\cite{mariconti2017mamadroid} (with 40-core CPU and 128G RAM), possibly because of 
our lower hardware configuration. 
Due to the dominating cost for tracing, {\tech} tends to incur higher 
total costs per app. However, the significantly higher sustainability and performance 
along with better resiliency may justify the reasonable extra costs.


\section{Threats to Validity and Limitations}\label{sec:discussion}
\vspace{-0pt}
In this section, we discuss various factors that may affect the 
validity of our results and performance of {\tech}, along with its 
limitations. 

One validity threat lies in the benchmark selection in our studies: 
the limited numbers of benign apps and malware used may not represent all respective Android apps. 
For a dynamic analysis, our dataset size is (at least among) the largest in the literature, but not 
as large as many used for static Android analysis approaches. 
To reduce this threat, we managed to obtain reasonably large datasets and purposely chose apps from 
a diverse set of sources. 

Results of dynamic analyses are commonly subject to the availability and coverage of run-time inputs, which 
poses another threat to our results (the characterization findings in particular). 
The traces used in our studies may not well represent all the behaviors of the apps. 
Profiling a large number of Android apps (which needs an effective automatic input generator) 
with high coverage is still an open research topic. 
To reduce this threat, we have used best resources available 
to obtain high-coverage inputs for our studies, at the cost of longer total time for using our 
technique (i.e., 10min tracing per app). 
However, for the characterization study, the most important findings were 
about the evolutionary patterns of run-time app behaviors (rather than the numbers themselves), which 
tend to be less sensitive to the input coverage. 
For our malware detection technique, the high and sustainable classification accuracy shown through 
extensive experiments suggest even lesser dependence on the coverage. 
Nevertheless, a minimal piece of app trace that is usable by our technique is required for classifying the app.

A threat to validity of our conclusion is that the characterization findings and 
the merits of {\tech} may not be exactly observed with different datasets. 
We attempted to reduce this risk through the diversity of our datasets. 
The general consistency in the results on {\base} between the authors' original study and ours on different datasets 
suggests that our study results may not be much biased, which helps support our conclusion drawn from the results. 

We have shown through a separate study (Study III) that our technique appeared to be resilient to 
encryption, reflection, and renaming attacks. 
{\tech} could be vulnerable to other advanced app obfuscation and malware evasion schemes. 
However, it only relies on identifying source/sink APIs from execution traces (instead of code). 
These APIs are part of the Android SDK classes, obfuscating which is not currently feasible. 
In addition, tt is well known that tricky malware can evade detection by withholding 
their malicious behaviors when they find being executed on an emulator. 
Our highly positive evaluation results obtained from traces collected on an emulator suggests that 
emulator evasion does not seem to much affect the effectiveness {\tech}. 
We have suggested that, due to the typically higher overhead of dynamic analysis (for profiling), 
{\tech} can take relatively longer time. 
Another limitation is that substantial updates in the Android SDK for source/sink APIs may 
affect the performance of our technique, and thus would trigger retraining 
(with the updated source/sink lists).

\section{Related Work}
\vspace{-0pt}
We discuss prior works closely related to ours, on {\em app characterization} and {\em malware detection} for Android.

\textbf{Characterization of Android apps.}
Previous characterization studies of Android apps mostly
focused on metrics extracted from the app code.
For example, in~\cite{Enck2011SAA}, 1,100 popular apps were studied to understand their use and misuse
of private information of mobile phones and users through reverse engineering.
A few dynamic characterization studies exist, which concern the installation and activation
methods of malware only~\cite{zhou2012dissecting} or execution
structure of benign apps only~\cite{cai2016understanding,Cai2017androidstudy}. 
Another examples include the dynamic study~\cite{wei2012profiledroid} that profiles
apps in terms of their inter-component communications (ICC) and network traffic.
CopperDroid~\cite{tam2015copperdroid} can be used for characterizing system calls in apps, and
the recently developed toolkit in~\cite{Cai2017droidfax} 
serves both static and dynamic characterizations.

In contrast, our characterization is purely dynamic, focusing on run-time sensitive accesses in both
benign and malicious apps.
More importantly, our characterization aims at a longitudinal examination of the
evolution of Android apps.
The characterization of evaluation datasets in~\cite{mariconti2017mamadroid} reveals the evolutionary
characteristics of both malware and benign apps also, but it instead focuses on API calls in apps and is
static.

\textbf{Android Malware Detection.}
Numerous approaches have been proposed for detecting Android malware,
by {\em statically} analyzing data and/or control flows~\cite{Feng14,grace2012riskranker,yang2014droidminer,Tripp2016PMM},
API use~\cite{wu2012droidmat,Aafer13,yang2014droidminer,zhang2014semantics,avdiienko2015mining}, and/or installation-time permission~\cite{Arp14,wu2012droidmat,saracino2016madam,avdiienko2015mining,Chen2016SSM}.
ICCDetector~\cite{xu2016iccdetector} distinguishes malware from benign apps based on their different patterns in ICCs.
Most recently, DroidSieve~\cite{suarez2017droidsieve} used resource-centric features to classify obfuscated apps,
while MamaDroid~\cite{mariconti2017mamadroid} aimed at sustainable detection, using a Markov model of API calls
after abstracting them to family and package levels.

Dynamic malware detectors have mainly exploited monitoring system and/or API calls in apps~\cite{burguera2011crowdroid,afonso2015identifying,Chen2016SSM}, or high-level behaviors exhibited through
the usage of system resources such as file and network access~\cite{shabtai2012andromaly}.
Some of these approaches used static features in addition to dynamic ones~\cite{lindorfer2015marvin,Chen2016SSM,saracino2016madam}.
Approaches relied on static calls in app code are generally vulnerable to code obfuscation (e.g., reflection)~\cite{suarez2017droidsieve}.

As suggested by the study in~\cite{mariconti2017mamadroid}, prior approaches are rarely shown to
be capable of sustainable detection. For example, DroidAPIMiner~\cite{Aafer13}, which was used as the baseline of
MamaDroid, did not even sustain its capabilities for a year without retraining~\cite{mariconti2017mamadroid}.
{\tech} targets a better sustainability (and resiliency) of malware detection than Mamadroid, using only
dynamic features based on the extent and distribution of {\em exercised} sensitive accesses and
vulnerable, method-level {\em control flows} in app executions. As a static detector, MudFlow~\cite{avdiienko2015mining}
builds a normal behavior model based on the statement-level sensitive {\em data flows} in a training set of benign apps in order to
identify new malware.

\section{Conclusion} 
As malware continues to be rampant in Android threatening its large user base,
defending against Android malware is crucial.
Numerous approaches have been proposed, mostly training a learning model to predict
the label of novel apps.
Yet, existing approaches tend not to sustain without retraining, which however is
not practical for detecting emerging malware.

In this paper, we introduce {\tech}, a novel malware detection approach based on a
new behavior profile of Android apps that models the distribution of their sensitive accesses.
We start with a longitudinal characterization of this profile.
Our findings reveal consistent differences between benign apps and malware over the past seven years,
despite the evolution of both groups. These findings explain why {\tech} sustains high accuracy for up to
at least four years, without retraining, as also supported by our extensive empirical evidences.
We also showed that {\tech} outperforms the state-of-the-art peer approach with demonstrated sustainability
in terms of the length of sustaining period for high accuracy and resilience to 
various obfuscation schemes, at reasonable costs.


\bibliographystyle{ACM-Reference-Format}
\bibliography{reference,security}



\end{document}